\documentclass[pra, twocolumn, aps, superscriptaddress, showpacs, groupaddress]{revtex4-1}
\usepackage{bigints}
\usepackage{amssymb}
\usepackage{graphicx}
\usepackage{amsmath}
\usepackage{dcolumn}
\usepackage{bm}
\usepackage{multirow}
\usepackage{hyperref}
\usepackage{physics}
\usepackage{bm}
\usepackage{comment}
\usepackage{xcolor}
\usepackage[caption=false]{subfig}
\usepackage{float}
\usepackage{siunitx}
\usepackage{dsfont}
\usepackage{bbold}
\usepackage{appendix}
\usepackage{relsize}
\usepackage{leftidx}
\usepackage{bbm}

\definecolor{lapislazuli}{rgb}{0.15, 0.38, 0.61}
\definecolor{YKblue}{rgb}{0.0, 0.18, 0.65}
\definecolor{carmine}{rgb}{0.81, 0.09, 0.13}
\definecolor{lavender}{rgb}{0.84, 0.79, 0.87}

\hypersetup{
	colorlinks=true,
	linkcolor=YKblue,
	linktoc=page,
	citecolor=blue,
	urlcolor=YKblue
}

\usepackage{tikz}
\usetikzlibrary{patterns,arrows,shapes,arrows,intersections,decorations}

\begin{document}
	
\title{Bose-Einstein condensation of photons in microcavity plasmas}

	\author{J. L. Figueiredo}
	\email{jose.luis.figueiredo@tecnico.ulisboa.pt} 
	
	\author{J. T. Mendon\c{c}a}
			
	\author{H. Ter\c{c}as}

	\affiliation{GoLP - Instituto de Plasmas e Fus\~{a}o Nuclear, Instituto
	Superior T\'{e}cnico, Universidade de Lisboa, 1049-001 Lisboa, Portugal}

\begin{abstract}
Bose--Einstein condensation of a finite number of photons propagating inside a plasma-filled microcavity is investigated. The nonzero chemical potential is provided by the electrons, which induces a finite photon mass allowing condensation to occur. We derive an equation that models the evolution of the photon-mode occupancies, with Compton scattering taken into account as the mechanism of thermalization. The kinetic evolution of the photon spectrum is solved numerically, and we find evidences of condensation for realistic plasma densities, $n_e\sim 10^{14} - 10^{15}\; \text{cm}^{-3}$, compatible with microplasma technology. The critical temperature is almost linear in the number of photons, and we find high condensate fractions at microcavity-plasma temperatures, for experimentally reasonable cavity lengths ($ 100-500 \; \mu$m) and photon numbers ($10^{10}-10^{12}$). 

\end{abstract}	
\maketitle

\textit{Introduction} $-$ Over the past years, Bose--Einstein condensation has been accomplished with atomic species, including $^{7}\text{Li}$ \cite{bradley}, spin-polarized $^{1}\text{H}$ \cite{fried}, metastable $^{4}\text{He}$ \cite{santos} and $^{41}\text{K}$ \cite{modugno}. Despite the remarkable advances on the experimental realization of BECs, the possibility of producing a condensate of photons remained elusive for a long time. The reason relies on the vanishing chemical potential of free photons, which leads to non-conservation of the number of particles during thermalization, thus preventing the second-order phase transition to take place. This problem was first circumvented in Ref.~\cite{chiao}, where it was shown that the presence of a cavity grants the photons with an effective mass. Nevertheless, no thermalization mechanism was proposed therein. The latter was then addressed in Refs.~\cite{klaers, klaersExp}, where experimental evidence for the formation of a photon BEC in dye-filled cavities were first reported. Later on, other authors have observed photon condensation in similar physical setups \cite{barlandExp, marelicExp}. Such remarkable findings have motivated a number of theoretical studies, unveiling the mechanisms behind photon condensation with dye molecules \cite{Klaers_2012, Sobyanin_2012, Snoke_2013, Kirton_2013, kruchkov} and atomic media \citep{kruchov_2013, Wang_2019}. \par

An alternative physical medium where one could imagine photons to undergo condensation is the plasma, where photon may also acquire an effective mass \cite{anderson, higgs, englert, guralnik}. Contrary to the case of the experiments with optical cavities, photon condensation in plasmas is thought to be a {\it bulk} phenomenon, arising in homogeneous and unbounded systems \cite{plasmaBEC}. This is particularly relevant in the astrophysical context, where external trapping potentials are absent. Indeed, the possibility of photon BEC in a plasma was first considered by Zel'dovich and Levich in 1968 \cite{zeldovich}, in relation to the distortion of the cosmic microwave background radiation through inverse Compton scattering $-$ the so-called Sunyaev-Zel'dovich effect \cite{sunyaev, birkinshaw}. However, the question of photon condensation in finite-sized plasma systems have never been proposed. \par

In this Letter, we investigate photon thermalization in a microcavity plasma, and find evidences of high-temperature condensation. The system under study takes advantage of both the photon mass, $m_{\rm ph} = \hbar \omega_p /c^2$ with $\omega_p$ the plasma frequency, and the boundary conditions induced by the cavity, which lead to a discretization of photon-momentum modes. The discreteness of the modes inside the cavity yields a finite critical temperature despite the system being effectively one-dimensional. We derive and solve the kinetic equations accounting for the evolution of the photon spectrum, with Compton scattering taken as the thermalization mechanism. After integrating out the electron degrees of freedom, we obtain a set of coupled equations for the photon modes dressed by the plasma. For sufficiently high photon intensities, we find macroscopic fractions of particles occupying the ground state. Moreover, the condensate energy can be varied by controlling the cavity length and electron density. Remarkably, the critical temperatures are extremely high, when compared to the ones of customary BEC experiments with identical photon numbers \cite{klaersExp}. We find that those temperatures are also compatible with plasma temperatures, which opens the possibility of conceiving condensation directly inside a microcavity plasma \cite{micCav1}. Since the microcavity can be easily built into a microelectronic circuit, the proposed mechanism finds a plethora of applications in a future generation of photon-based devices.\par
\textit{Plasma wave equations.}--- We start by revising the theory of electron-photon coupling in a plasma. Essentially, the effect of the plasma is to modify the refraction index of the medium, which becomes $n(\omega) = (1-\omega_p^2/\omega^2)^{1/2}$, with $\omega_p^2 = e^2 n_e^2/\epsilon_0m_e$ and $e$ being the elementary charge, $n_e$ the electron density, $\epsilon_0$ the vacuum permittivity, and $m_e$ the electron mass.  The frequency becomes space and time dependent, through the local electron density $n_e \equiv n_e(\mathbf r,t)$. Conversely, the ion motion is negligible due their high inertia, and the photon dynamics is mostly determined by the electrons. The photon dispersion relation follows from $\omega = ck/n(\omega)$ \cite{mendonca2000theory}. \par 
To derive the details of the photon-plasma coupling, we resort to Maxwell equations. We start from Amp\`ere's law
\begin{equation}
{\bm \nabla} \times {\bm{\mathcal B}}=\frac{1}{c^2}\partial_t \bm{\mathcal E} + \mu_0\bm{ J}, \label{eq_photons}
\end{equation}
with $\bm{\mathcal B}$ denoting the magnetic field, $\bm{\mathcal E}$ the electric field and $\bm{ J}$ the charge current density. The latter is responsible for the coupling of photons with the plasma via $\bm{ J} = \sum_{j} Q_jn_j \mathbf u_j$, with $j=\{e,i\}$ running over the different species (electron, $e$,  and ion, $i$, for definiteness) of charge $Q_j$, density $n_j$ and velocity $\mathbf u_j$. The fields $n_j$ and $\mathbf u_j$ evolve with their own classical equations of motion coupled to the electromagnetic fields,
\begin{align}
	&\partial_t n_j + \bm{\nabla}\cdot (n_j \mathbf u_j) = 0 \ \ \text{and} \label{dens}\\
	&\partial_t \mathbf u_j + \mathbf u_j \cdot \bm \nabla \mathbf u_j = \frac{Q_j}{m_j} (\bm{\mathcal E} + \mathbf u_j \times \bm{\mathcal B}) - \frac{1}{m_jn_j} \bm{\nabla} P_j, \label{vel}
\end{align}
where $m_j$ is the mass of the specie $j$ and $P_j$ is the pressor tensor. Following the usual prescription, we write the electromagnetic fields in terms of the potentials $\phi$ and $\mathbf A$, $\bm{\mathcal E} = - \bm \nabla \phi - \partial_t \mathbf A $ and $\bm{\mathcal B} = \bm \nabla \times \mathbf A$. Replacing for those in Eq.~\eqref{eq_photons} and neglecting the slow ion motion leads to
\begin{equation}
	\Big(\bm \nabla^2 - \frac{1}{c^2}\partial_t^2\Big) \mathbf A= \frac{\omega_p^2}{c^2} \mathbf A. \label{KG}
\end{equation}
Equation \eqref{KG} takes the form of a Klein--Gordon equation, which is a consequence of the photons acquiring a mass, in a process that is reminiscent to the Higgs-Anderson mechanism \cite{anderson, higgs, englert, guralnik}. By Fourier transforming Eq. \eqref{KG}, we obtain the photon dispersion
\begin{equation}
	\omega \equiv \omega_\mathbf{k} = \big(\omega_p^2 + c^2 k^2\big)^{1/2}, \label{dispRel}
\end{equation}
which, when compared to the relativistic formula for the energy, leads to the photon mass $m_{\rm ph} = \hbar \omega_p/c^2$, scaling with the electron density as $m_{\rm ph}\sim \sqrt{n_e}$. Equation \eqref{dispRel} can now be compared to $\omega = ck / n$, from which we can extract the refraction index $n(\omega) = (1-\omega_p^2/\omega^2)^{1/2}$.\par
\textit{Kinetic model}.--- In the case of a fully ionized plasma, elastic electron-photon scattering is the main source of thermalization. We follow the Boltzmann approach and calculate the variation of the number of particles measured by a joint distribution function $\rho(\mathbf p, \mathbf k,t)$ with electrons in mode $\mathbf p$ and photons in mode $\mathbf k$, at time $t$. We have
\begin{equation}
	\partial_t \rho(\mathbf p, \mathbf k,t) = J_{+}(\mathbf p, \mathbf k,t) - J_{-}(\mathbf p, \mathbf k,t),
\end{equation}
with $J_{+(-)}$ being the number of particles per unit volume per unit time that enters (leaves) the phase-space element $\dd^3\mathbf p \; \dd^3\mathbf k$ centered in $(\mathbf p,\mathbf k)$ due to a scattering event. The currents can be written as
\begin{align}
&J_{+}(\mathbf p, \mathbf k,t) = \int \dd^3\mathbf p'\dd^3\mathbf k' \rho(\mathbf p',\mathbf k',t) w(p',k' \rightarrow p,k) \nonumber \\
& \quad \quad \quad \quad \quad \quad \quad \quad \times [1+N(\mathbf k,t)][1-F(\mathbf p,t)] ,\label{Jin} \\
&J_{-}(\mathbf p, \mathbf k,t) = \int \dd^3\mathbf p'\dd^3\mathbf k' \rho(\mathbf p,\mathbf k,t) w(p,k \rightarrow p',k') \nonumber \\
& \quad \quad \quad \quad \quad \quad \quad \quad \times [1+N(\mathbf k',t)][1-F(\mathbf p',t)], \label{Jout}
\end{align}
with $N(\mathbf k,t) = n_e^{-1} \int \dd^3\mathbf p \ \rho(\mathbf p, \mathbf k,t)$ and $F(\mathbf p,t) = n_{\rm ph}^{-1} \int \dd^3\mathbf k \ \rho(\mathbf p, \mathbf k,t)$ denoting the photon and electron distributions with densities $n_{\rm ph}$ and $n_e$, and total number of particles $N_{\rm ph}$ and $N_{\rm e}$, respectively. The factor $w(p,k\rightarrow p',k')$ is the transition rate from incoming $(p,k)$ to final $(p',k')$ states, with $p,\;k,\;p'$ and $k'$ the four-vector momenta associated with the electron and photon degrees of freedom. Additionally, the factors $1+N$ and $1-F$ in Eqs.~\eqref{Jin} and \eqref{Jout} account for quantum degeneracy of each population, i.e., they ensure that fermions do not occupy the same state and bosons tend to occupy the same state. For the conditions considered here (micro-discharge plasmas), electrons are non-degenerated and we may set $1-F \simeq 1$. On the contrary, the photon degeneracy may not be discarded, for it results in a non-linear term of order $N^2$ that is crucial to the condensation process. \par 

As we are interested in the dynamics of photons, it is convenient to integrate out the electron degrees of freedom. This procedure is valid as long as the correlations between electrons and photons can be neglected; in other words, when the following expansion holds,
\begin{equation}
	\rho(\mathbf p,\mathbf k,t) \simeq F(\mathbf p,t)N(\mathbf k,t) + \text{correlations},
\end{equation} 
with the second term on the right hand side being much smaller than the first. In fact, correlations are small whenever there is a separation of time scales, which in the present case amounts to have the electron gas equilibrated much faster than the photons. Under those assumptions, and invoking dynamical reversibility in the form $w(p,k\rightarrow p',k') = w(p',k'\rightarrow p,k)$, the Boltzmann equation reduces to an equation for the photon distribution function,
\begin{align}
	&\partial_t N(\mathbf k) = \frac{1}{n_e}\int \dd^3\mathbf p \; \dd^3\mathbf p'\;\dd^3\mathbf k' \ w(p,k \rightarrow p',k')  \nonumber \\
	 &\!\!\!\times \big\{ F(\mathbf p')  N(\mathbf k')[1\!+\!N(\mathbf k)]-F(\mathbf p)N(\mathbf k)[1\!+\!N(\mathbf k')]\big\}.\label{Boltzmann2}
\end{align}
The amplitude of the Compton scattering is given by
\begin{equation}
	w = \frac{3\sigma_Tn_e}{16\pi}\delta(p+k - p - k')\big(1+\cos^2\theta\big), \label{ComptonCS}
\end{equation}
with $\sigma_T\simeq 6.65 \times 10^{-29} \ \! \text{m}^2$ being the Thompson cross-section and $\theta$ the photon scattering angle.\par 
 At this point, it is convenient to introduce the discretized photon momenta $\mathbf k \equiv \mathbf{k}_\ell = \pi \ell \mathbf{e}_z/d$, where $\ell$ is an integer, $d$ is the cavity length and $\mathbf e_z$ is directed along the longitudinal axis of the cavity. Such discretization is achieved for a cavity of planar mirrors separated by a distance $d$ and width $w\ll d$. The discretized photon frequencies will be denoted by $\omega_\ell \equiv \omega_{\mathbf{k}_\ell}$. We will further suppose that the photons are confined to move in one dimension along the cavity axis, which results in recasting the Compton amplitude as 
\begin{equation}
	\int \dd^3 \mathbf{k}' \ w(p,k \rightarrow p',k') \rightarrow \sum_{\ell '} \tilde w(p,k_\ell \rightarrow p',k_{\ell'}), \label{appx}
\end{equation}
where $\tilde w$ is the appropriate transition rate in terms of the discretized photon momentum \cite{Note1}. Deviations from the $z-$direction are suppressed due to the $\cos^2\theta$ dependence in Eq.~\eqref{ComptonCS}. \par
Assuming thermal equilibrium for the plasma, the electron distribution can be approximated by a Maxwell-Boltzmann function at temperature $T_e$, $F(\mathbf k) = F_0\text{exp}(-E_{\mathbf k}/k_BT_e)$, where $E_{\mathbf k} =\hbar^2k^2/(2m_e)$ is the electron dispersion and $F_0$ ensures the normalization $\int \dd^3 \mathbf k \ F(\mathbf k) = n_e$. The electron equilibrium is assumed to be maintained throughout the experiment, such that $\partial_t F \simeq 0$ is valid during the photon equilibration process. The Boltzmann equation can then be simplified to the following balance equation
\begin{equation}
	\partial_{\tau} N_\ell = \sum_{\ell'} \big[N_{
	\ell'}\mathcal W_{\ell \ell'} - N_{\ell} - (1-\mathcal W_{\ell \ell'})N_{\ell}N_{\ell'}\big], \label{cav1}
\end{equation}
with $N_\ell \equiv N(\mathbf k_\ell,\tau)$ the photon-mode occupancies. Here, $\Delta_{\ell\ell'} = \hbar(\omega_{\ell'} -\omega_\ell)/(k_BT_e)$ are normalized energy shifts, such that $\mathcal W_{\ell\ell'}=\text{exp}(\Delta_{\ell\ell'})$. Moreover, $\tau=t/t_0$ is the dimensionless time given in units of the typical thermalization timescale $t_0 = 8\pi/(\sigma_T n_e c)$. Notice that the total photon energy and entropy are not conserved quantities \cite{Rev}. This happens since the photons are formally in an open system with the plasma acting as a reservoir. Therefore, no $H$--theorem exists for the particular set of Eqs.~\eqref{cav1}. The latter would be verified only if the full plasma kinetics had been taken into account. \par 
The last term on the right-hand side of Eq.~\eqref{cav1} vanishes as $T_e\rightarrow \infty$, and the equation becomes linear in $N_\ell$, which prevents the formation of a condensate \cite{Levermore,Josserand}. Hence, there must be a critical temperature $T_c$ above which the condensate no longer develops. In cold atom experiments, the value of $T_c$ typically ranges between a few nK and $\mu$K, depending on the density and mass of the atomic cloud \cite{Anglin_2002}. In fact, theoretical calculations reveal that $T_c$ scales as $T_c \sim 1/m$, with $m$ typically in the range $10^{-27}-10^{-26}$ kg for atomic BECs. In the present case, an estimate for the photon mass is given by $\hbar \omega_p/c^2$, which is about 12 to 13 orders of magnitude below the typical atomic masses. We can thus anticipate much higher critical temperatures.\par

\begin{figure}[t!]
\centering 
\hspace*{-17pt}
\includegraphics[scale=0.55]{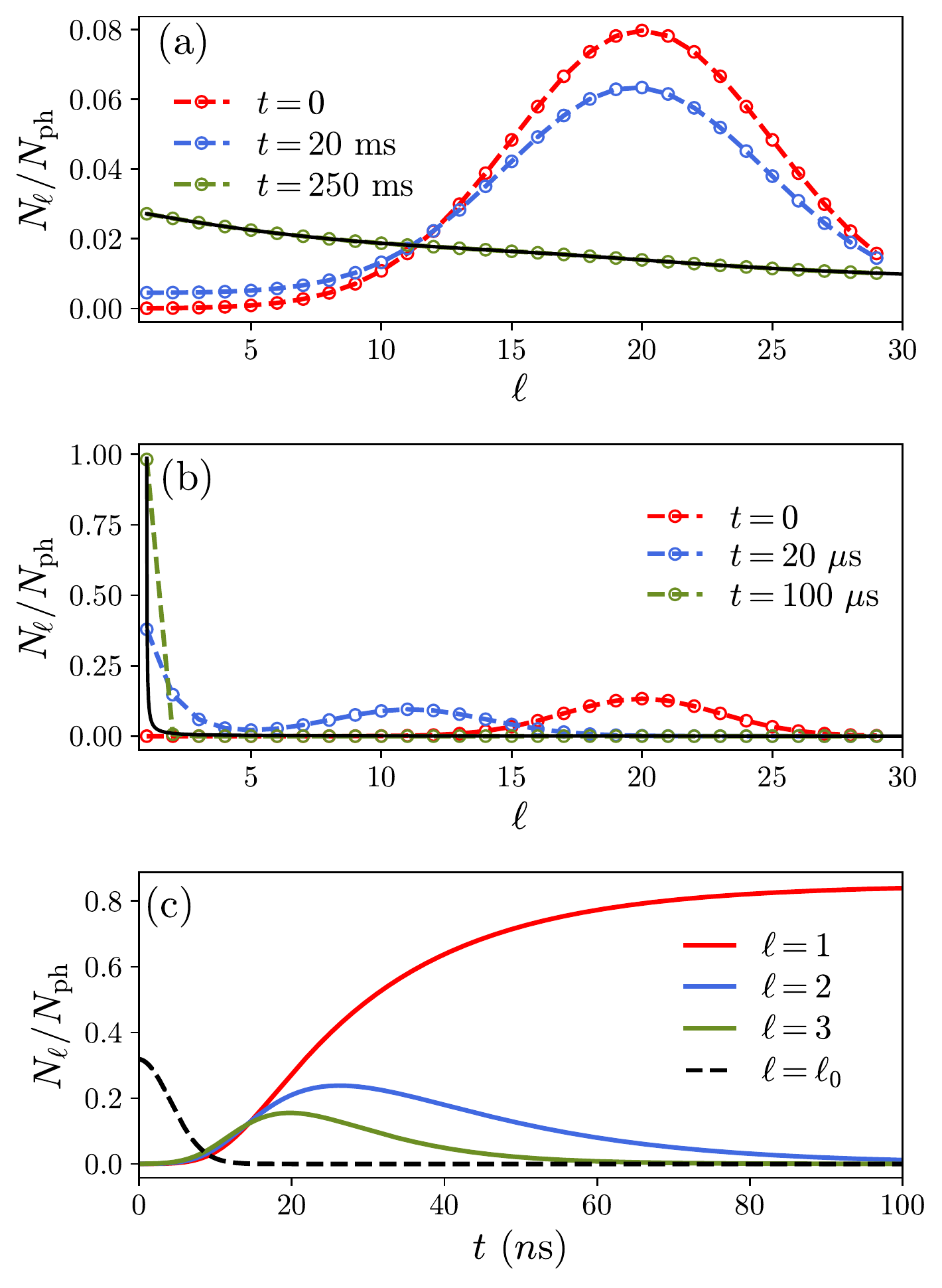}
\caption{Photon spectra for the two distinct regimes for $N_\text{ph}=10^8$: (a) $T_e = 2\times 10^5 \ \!$eV ($T_e> T_c$) with steady-state Bose-Einstein distribution; (b) $T_e = 3 \ \! $eV ($T_e< T_c$) with the formation of a condensate. The solid black lines shows the Bose-Einstein distribution at the plasma temperature, after the system had reached thermal equilibrium. (c) Dynamical profiles of the mode occupancies displaying condensation as a function of time, with $N_\text{ph}=10^{11}$ and $T = 3\;$eV. Other parameters are $d = 100\; \mu$m, $n_e = 10^{14} \; \text{cm}^{-3}$, $\ell_0 = 20$, and $\Gamma = 5$.}
\label{fig1}
\end{figure}

\textit{Thermalization and condensation.}--- 
 Solutions to Eq.~\eqref{cav1} have been obtained numerically, using a fourth order Runge-Kutta method. The occupancies were initiated with a Lorentzian distribution centered at $\ell = \ell_0$, with bandwidth $\Gamma$ and total number of photons $N_\text{ph}$, 
\begin{equation}
	N_\ell(0) = \frac{N_\text{ph}}{\pi} \frac{\Gamma/2}{(\ell - \ell_0)^2 + (\Gamma/2)^2}. \label{initEq}
\end{equation}
The solutions depend on the electron temperature, total number of photons, cavity length and plasma frequency. Figure \ref{fig1} [panels (a) and (b)] shows the initial, intermediate and steady-state occupancies as a function of the mode number, for the case of thermal and condensate steady-states (we illustrate the case $\ell>0$ since the steady-state solution is symmetric with respect to the sign of $\ell$). 
In steady-state, the photons tend to a Bose-Einstein distribution,
\begin{equation}
	f(\varepsilon,T,\mu) = \frac{1}{\text{exp}(\frac{\varepsilon - \mu}{k_B T})-1}, \label{BE}
\end{equation}
with $T = T_e$ signaling that photons thermalize with the plasma electrons. 
The condensation time is of the order of tens of nanoseconds, which is much smaller than the time of formation of atomic BECs via evaporative cooling \cite{Chaudhuri_2007}. \par 
\begin{figure}
\centering 
\hspace*{-13pt}
\includegraphics[scale=0.51]{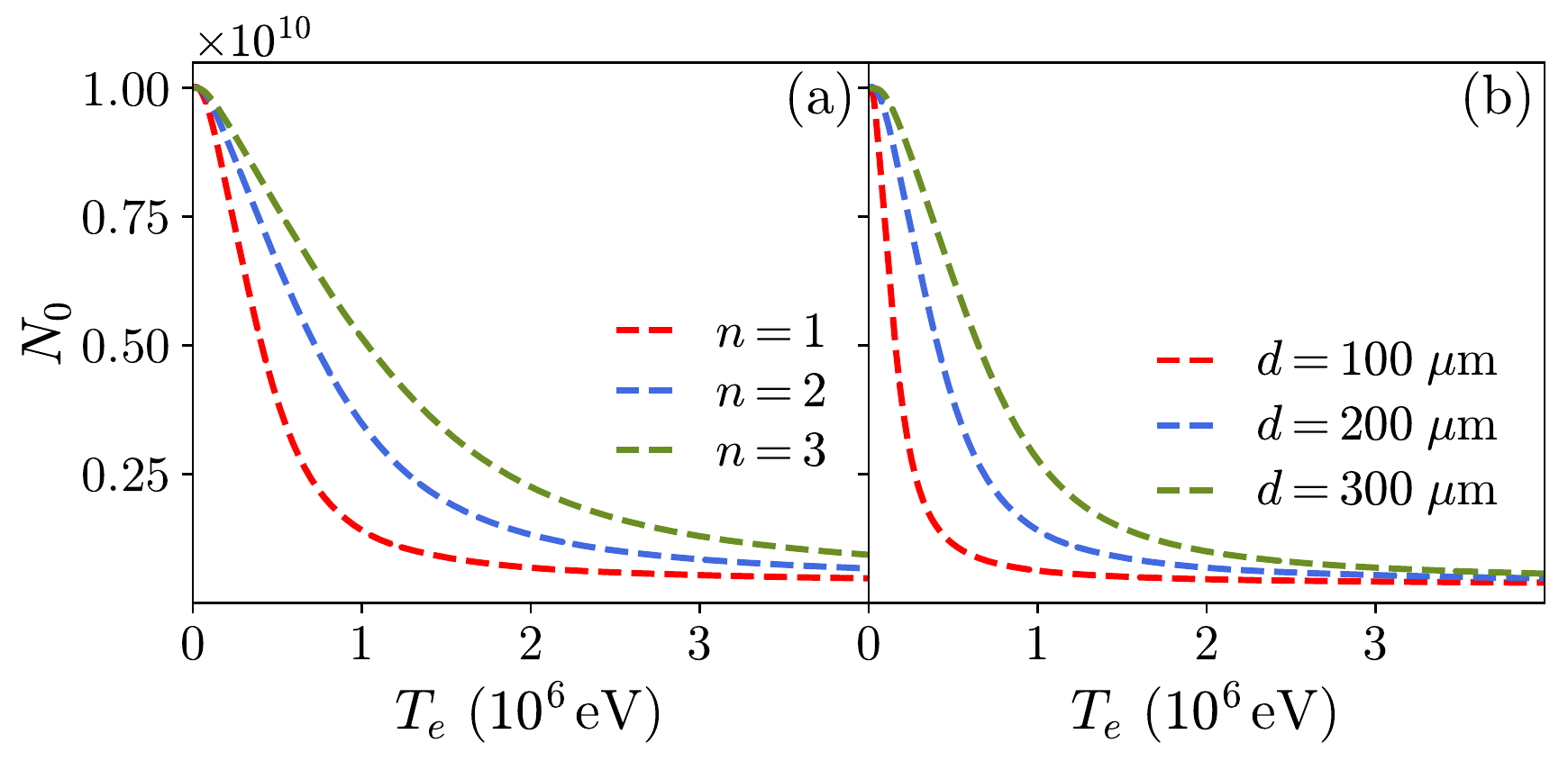}
\caption{Condensate fraction as a function of the electron temperature, (a) for different values of the reduced photon number $n = N/N_{\rm ph}$ and cavity length, and (b) for different values of the cavity length with $N_\text{ph}=10^{12}$.}
\label{fig2}
\end{figure}

\begin{figure}
\centering 
\hspace*{-11pt}
\includegraphics[scale=0.52]{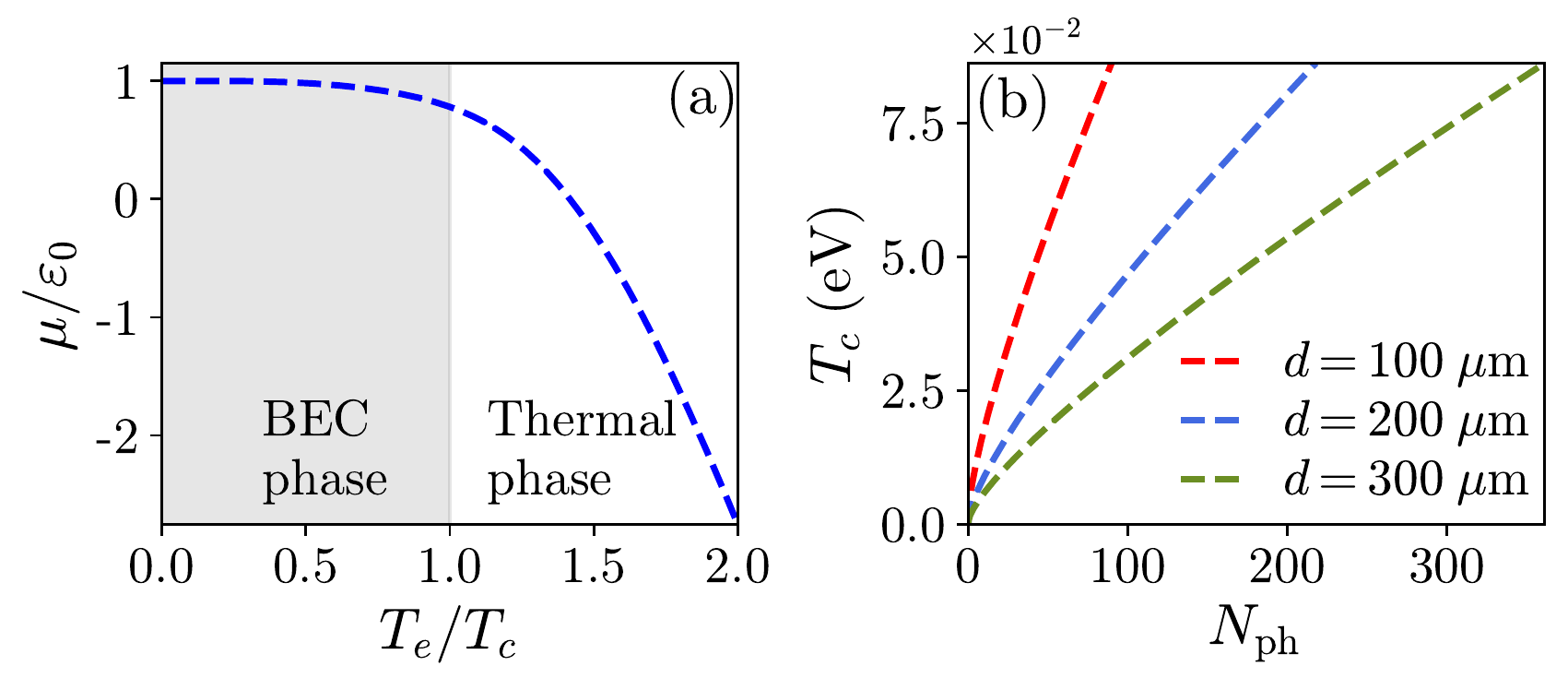}
\caption{(a) Typical chemical potential normalized profile as a function of the electron temperature. The BEC phase is the region of almost constant $\mu$; when $\mu$ decreases, the system enters in the thermal phase, with distribution spectrum of a thermal Bose gas. (b) Critical temperature as a function of the total number of photons for different cavities and  $n_e = 10^{14} \; \text{cm}^{-3}$. The critical temperature approaches a straight line for $d\gg c/\omega_p$.}
\label{fig3}
\end{figure}

The crossover between the BEC and thermal phases is governed by the chemical potential, which is fixed by the temperature and total numbers of particles through $N_\text{ph} = \sum_\ell f( \hbar \omega_\ell,T,\mu)$. The latter bears a solution of the form $\mu \equiv \mu(N_\text{ph},T)$. When the number of photons surpasses a critical number $N_c$, the excess particles occupy the ground state, which is possible only if $\mu(N_\text{ph}>N_c,T) \sim \varepsilon_0$, with $\varepsilon_0$ the ground-state energy, so that Eq.~\eqref{BE} attains large values at the origin. In Fig.~\ref{fig2}, we depict the condensate fraction as a function of $T_e$. The chemical potential was also determined numerically and the result is shown in the left panel of Fig.~\ref{fig3}.  \par
It is also convenient to obtain an analytical estimate for $T_c$. The exact definition requires separating the contribution of the number of particles in the ground-state from the remaining states, $N_\text{ph}-N_0 = \sum_{\ell\neq \pm 1} f(\hbar\omega_\ell,T,\mu)$. At the critical temperature, we replace $\mu$ by $\varepsilon_0$ and neglect $N_0$, to get
\begin{equation}
	N_{\rm ph} = g \sum_{\ell = 2}^{\infty} f(\hbar\omega_\ell,T_c,\varepsilon_0), \label{critic}
\end{equation}
where $g=2$ is the degeneracy factor. Typically, an analytical estimate for $T_c$ is available in the thermodynamic limit (in this case, that is $d\rightarrow \infty$ and $N_\text{ph}\rightarrow \infty$ while $N_\text{ph}/d$ is maintained finite). However, as it has been recognized, the thermodynamic limit yields $T_c = 0$ when the spacial dimension of the condensate is less than three \cite{1DBec}. Although this prevents condensation from developing in very large systems, the result is modified when the system is considered finite. Therefore, instead of taking the thermodynamic limit, we simply assume that $d\gg c/\omega_p$ while being finite. As long as the energy spacing is negligible compared to the temperature, the summation in Eq.~\eqref{critic} can be replaced by an integral, and we obtain
\begin{equation}
	T_c \approx \frac{\hbar^2 k_{0}^2}{\xi  m_{\rm ph} k_B} N_\text{ph}, \label{TcApprox}
\end{equation}
with $k_0=\pi/d$ the ground-state wavevector and $\xi = 2\pi - 4 \arctan{2}\simeq 1.9$ a constant. The rigorous relation is obtained by evaluating Eq.~\eqref{critic} numerically, which we show in panel (b) of Fig.~\ref{fig3} for microcavity lengths. \par 
As anticipated above, the value of $T_c$ is much higher than the typical values of atomic BECs, stemming from the small photon mass. That can be deduced analytically since, for large $d$, the photon dispersion approaches $\varepsilon_k\simeq  m_{\rm ph} c^2 + \frac{\hbar^2 k^2}{2m_{\rm ph}}$, which is quadratic, akin to massive condensation of Schr\"{o}dinger bosons \cite{Rev}. Additionally, Eq.~\eqref{TcApprox} gives $T_c = 0$ when $d\rightarrow \infty$ and $N_\text{ph}/d$ is finite, due to the dependence on $N_\text{ph}/d^2$. This is consistent with previous investigations on finite-sized BECs \cite{Ketterle_2}.\par

\textit{Conclusions.}--- We derived a kinetic model for the evolution of one-dimensional photon modes in contact with a plasma, starting from the Boltzmann equation. The electron population is considered to be in constant equilibrium at temperature $T_e$, which modifies the photon dispersion (by opening a gap of $\hbar \omega_p$ at $k=0$) and thermalizes the photon gas due to multiple Compton scattering. After integrating out the electron degrees of freedom, we obtained an effective equation for the photons which resembles a balance equation of statistical physics, that we solved numerically. The solution showed that the photon gas approaches a Bose-Einstein distribution at the plasma temperature, which admits a finite-sized condensed phase for sufficiently small temperatures. These temperatures are, however, much higher than the typical values for customary BEC, and can easily surpass the room temperature for moderate photon numbers. The reason lies on the much smaller photon masses for the present configuration, $m_{\rm ph} \simeq 10^{-40}\;\!$kg. \par 
Experimental implementation of photon condensation as described here requires resonators with a high quality factor $Q$, with the latter being defined as the ratio of initial to loss energy per oscillation cycle. The typical time of condensation ranges from the nanosecond to the microsecond time scale depending on the total number of photons. By estimating the photon lifetime inside the cavity we obtain restrictions on the quality factor that are compatible with the existing state-of-the-art technology (see \cite{SM}). A microplasma discharge can be used to obtain a homogenous plasma of about $\sim 100-500 \; \mu$m size as required for the thermalization \cite{microplasma}. Moreover, the plasma must not touch the mirrors and thus should be contained inside a transparent cell. The latter may induce quantitative corrections to the quality factor, which must be compensated by the intensity of the laser. We refer to the Supplemental Material for additional experimental details \cite{SM}.  \par 
The present work differs from the conventional case of photon condensation in dye-filled microcavities \cite{klaersExp,barlandExp,marelicExp}, where the photon mass is determined by the cavity cut-off frequency, typically in the range of $10^{14}$ Hz. Here, the plasma frequency establishes even smaller photon masses, resulting in higher critical temperatures. On the one hand, the ground-state wavelength is determined by the cavity distance, which gives an extra degree of control over the BEC parameters. This opens the possibility of setting the BEC wavelength over a wide range of values, which may have technological applications on the search for new light sources. On the other hand, the microplasma technology can be adapted to fit inside small circuits, which may allow direct implementation of the condensate in a future generation of photon-based devices. In particular, the condensate constitutes a potential candidate for a quantum battery \cite{QuantBAt,QuantBAt2}. Extensions to the case of solid-state degenerate plasmas $-$ eventually leading to condensation of photons in a partially filled cavity $-$, as well as the inclusion of the reservoir dynamics, deserve further investigation.\par

\textit{Acknowledgments.}--- J. L. F. and H. T. acknowledge Funda\c{c}\~{a}o da Ci\^{e}ncia e a Tecnologia (FCT-Portugal) through the Grants No. PD/BD/135211/2017, UI/BD/151557/2021, and through Contract No. CEECIND/00401/2018 and Project No. PTDC/FIS-OUT/3882/2020, respectively. The authors also acknowledge the Referees for their important queries, which helped to ameliorate the manuscript.


\twocolumngrid
\bibliography{references.bib}

\end{document}


\title{Supplemental Material on ``Bose-Einstein condensation of photons in microcavity plasmas"}

\author{J. L. Figueiredo}
\affiliation{\it Instituto de Plasmas e Fus\~{a}o Nuclear, Instituto Superior T\'{e}cnico,
Universidade de Lisboa, 1049-001 Lisbon, Portugal}

\author{J. T. Mendon\c{c}a}
\affiliation{\it Instituto de Plasmas e Fus\~{a}o Nuclear, Instituto Superior T\'{e}cnico,
Universidade de Lisboa, 1049-001 Lisbon, Portugal}

\author{H. Ter\c{c}as}
\affiliation{\it Instituto de Plasmas e Fus\~{a}o Nuclear, Instituto Superior T\'{e}cnico,
Universidade de Lisboa, 1049-001 Lisbon, Portugal}

\begin{abstract}
In this Supplemental Material, we provide experimental details on the microcavity plasma configuration required to achieve photon condensation. We estimate the rate of laser absorption by inverse bremsstrahlung and show that, for the cavity and plasma parameters considered here, a sufficient number of photons can be stored inside the cavity with a moderate laser intensity in order that absorption becomes negligible. 
\end{abstract}
\maketitle

\tableofcontents 

\section{Experimental requirements and set-up}
Microcavity plasmas are low-temperature plasmas created inside a cavity with at least one mesoscopic dimension ($\sim 1-100 \;\mu$m). Those are generated by applying a voltage to a microcavity filled with a gas, so that an electric current passes through and produces the plasma through ionization. Electronic densities in the range $10^{13}-10^{16} \; \text{cm}^{-3}$ and temperatures of about $1-5$ eV are typical values for this technology. Microplasmas find a wide range of applications in biology \cite{biol}, spectroscopy \cite{spet}, radiation sources \cite{radS} and material synthesis \cite{MaterialS}. See \cite{REV} for a review on the topic. \par
Since photon condensation in a microcavity depends drastically on the electronic density, temperature and cavity distance, we find the ideal parameters to be those summarized in table~\ref{tab1}, which are all compatible with current microplasma technology. 
\begin{table}[h]
\centering
\begin{tabular}{ | ccccccc | ccccccc | } 
\hline
&&&$T_e$ & && & &&&$3 \; \text{eV}$&&& \\ 
\hline 
&&&$n_e$ & &&& &&&$10^{14} \;\text{cm}^{-3}$&&& \\
\hline 
&&&$d$ &&&&   &&&$100\; \mu m$&&&\\
 \hline 
\end{tabular}
\caption{Ideal plasma and microcavity parameters based on standard values.}\label{tab1}
\end{table}
Simulations of the photon evolution equations using the parameters of table~\ref{tab1} are shown in figure~(1) of the main text. We find condensation times that range from $\tau_\text{BEC} = 100\;$ns ($N_\text{ph} = 10^{11}$) to $\tau_\text{BEC} = 10^6 \;$ns ($N_\text{ph} = 10^8$). We will see in the following section that a sufficiently small condensation time is required in order to render laser absorption by the plasma negligible within the time of condensate formation. Calculations regarding the laser absorption coefficient indicate that $N_\text{ph}$ should be of at least $10^{11}$ (see the following section). \par
On the other hand, the validity of equation~(13) of the main text relies on the assumption that the number of electrons largely surpasses the number of photons, in order for the plasma to be regarded as a thermal reservoir. By estimating the number of electrons by $N_e = n_e Ad$, with $A$ the cross-sectional area of the cavity, we obtain $N_e \sim 10^{14}$ for $A\sim \text{tens of} \ \text{cm}^2$, which fits the requirement. An additional source of losses are the leaks through the mirrors, which will always happen for sufficiently long storage time. Therefore, we must also assure that the condensate has time to form before those losses become significant. The time scale for mirror losses can be estimated as $\tau_d = 2Qd/c$, where $Q$ is the microcavity quality factor. The condition $\tau_\text{BEC}\ll \tau_d$ sets a minimum scale for $Q$ 
\begin{equation}
Q \gg 10^4 ,
\end{equation}
which is easily achieved experimentally \cite{Q1,Q2}. Finally, we must also guarantee that the time it takes for a laser of moderate intensity to fill the cavity with the required number of photons is much smaller than the condensation time. We note that the intensity can not be increased indefinitely since we must avoid material damage. We can assess the time of filling of the cavity using $\tau_\text{L} = N_\text{ph} \hbar \omega_\text{L}/(A I)$, where $I$ is the laser intensity and $\omega_\text{L} = 2\pi \times  3\times 10^{13}\;$Hz is the laser frequency used in the simulations of the main text. Setting $\tau_\text{L} \ll \tau_\text{BEC}$ results in a condition for $I$
\begin{equation}
	I \gg \frac{N_\text{ph}\hbar\omega_\text{L}}{\tau_\text{BEC}A} \sim  1\text{W}/\text{cm}^2.
\end{equation}
We conclude that a moderate intensity laser suffices to achieve the desired result.\\

\begin{figure}[H]
\centering 
\includegraphics[scale=0.33]{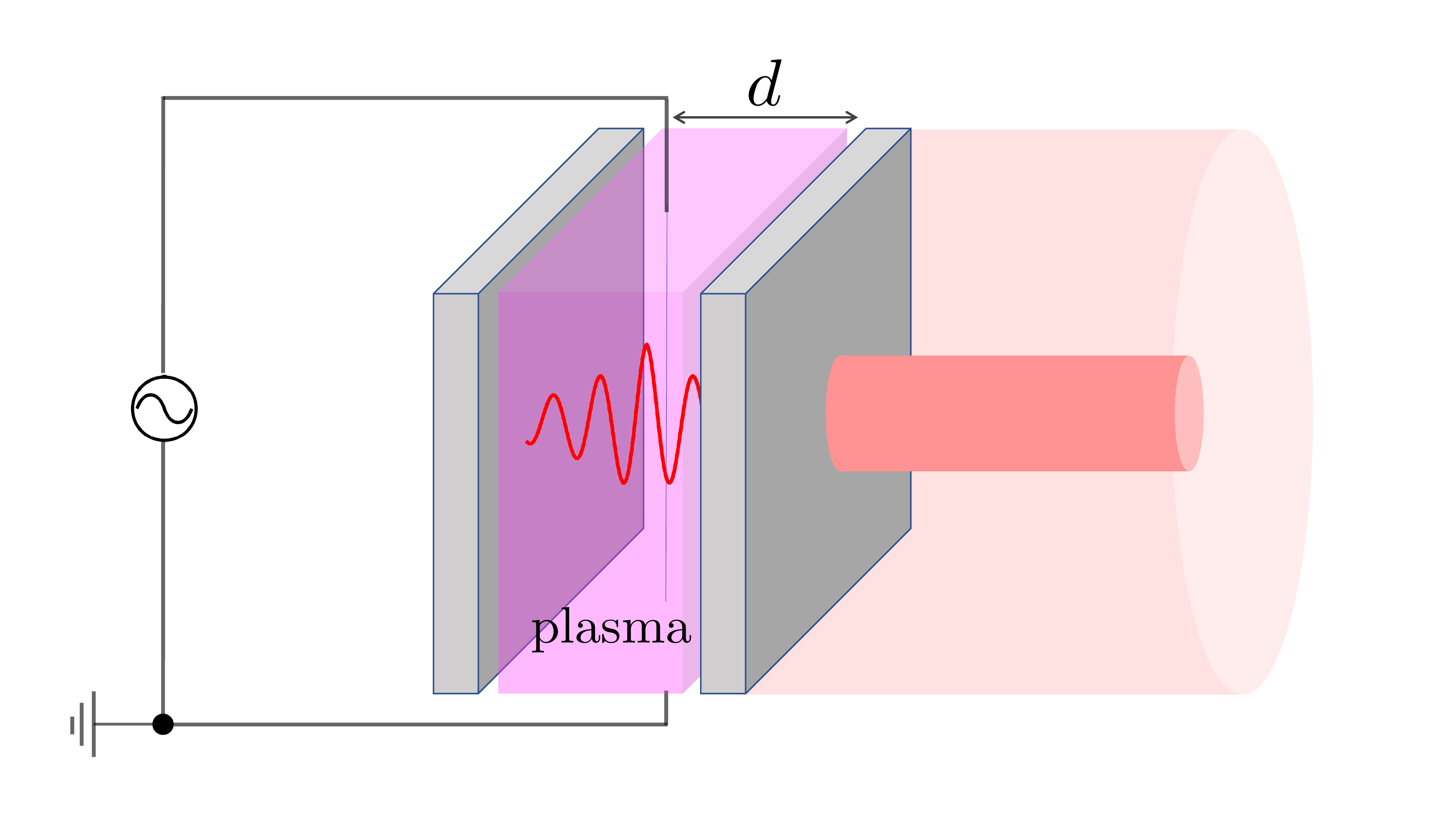}
\caption{Scheme of the experimental set-up aimed at photon condensation inside a plasma microcavity. The cavity is composed by parallel mirrors separated by a distance $d$, with quality factor $Q$ and cross-sectional area $A$ that can trap the radiation inside during a time $\tau_d$. A gas is placed inside the cavity and a microplasma is produced after an electrical discharge hits the gas. The plasma density is controlled by the applied current. A laser of frequency $\omega_\text{L}$ is shined perpendicular to the cavity mirrors during an interval $\tau_\text{L}$, so that a total number of $N_\text{ph}$ photons is stored inside.}
\label{fig1}
\end{figure}

The hierarchy of experimental conditions given in this section can be summarized as follows 
\begin{equation}
	\tau_\text{L} \ll \tau_\text{BEC} \ll \tau_\text{d}.
\end{equation}
As stated before, laser absorption by inverse bremsstrahlung must also be taken into account, and a more detailed analysis is given in the next section. We also show a possible experimental set-up in figure~\ref{fig1}.

\section{Laser absorption by inverse bremsstrahlung} 

Absorption by inverse bremsstrahlung is the most efficient source of heating in laser-driven plasmas. It is a type of collisional absorption that happens when an electron, trapped in the electrostatic field of an ion, absorbs a photon and accelerates. The photon energy is transferred to the plasma in such a way that the total number of photons is not conserved, which contributes to the diminishing of the laser while it travels in the medium. Since electron-ion collisions are also present in the equilibrium state of the microplasma, it is important that we make sure the absorption is vanishingly small in the devised conditions. Strong absorption of the condensed photons would necessarily spoil the condensation phenomena.  \par
To estimate the absorption rate, we calculate the absorption coefficient $\alpha_\text{abs}$, defined as
\begin{equation}
\alpha_\text{abs} = \frac{I_\text{in}-I_\text{out}}{I_\text{in}}	,
\end{equation}
where $I_\text{in}$ and $I_\text{out}$ are the incoming and outgoing laser intensities, respectively. The change in the laser intensity is governed by 
\begin{equation}
	\frac{dI}{dz} = -\kappa I ,
\end{equation}
where $\kappa$ is the spatial damping rate of the laser due to inverse bremsstrahlung. The absorption coefficient per half round trip can be decuced,
\begin{equation}
	\alpha_\text{abs} = 1- \text{exp}\Big(-\int_{0}^{d} dz \ \kappa \Big) , \label{alpha}
\end{equation}
where $d$ is the cavity length. For the case being considered here, the plasma is sufficiently homogeneous so that we can approximate $\alpha_\text{abs} \approx 1-e^{-\kappa d}$. The damping rate for frequency $\omega$ can be calculated as
\begin{equation}
	\kappa (\omega) = \frac{\nu_\text{ei}}{c}\frac{\omega_p^2}{\omega^2}\Bigg(1-\frac{\omega_p^2}{\omega^2}\Bigg)^{\!\!-1/2}, \label{ki}
\end{equation}
where $\omega_p$ is the plasma frequency, $c$ is the speed of light and $\nu_\text{ei}$ is the electron-ion collision frequency responsible for the inverse bremsstrahlung process  \cite{Eliezer}. We can immediately see that when $\omega$ is arbitrarily close to $\omega_p$, $\kappa \rightarrow \infty$ and consequently $\alpha_\text{abs} \approx 1$, so that absorption is highly effective. To avoid absorption, the ground-state frequency, defined as 
\begin{equation}
\omega_0 = \left(\omega_p^2 + \frac{c^2 \pi^2}{d^2}\right)^{1/2},
\end{equation}
must surpass $\omega_p$ by a large amount. This is accomplished when $d$ is of the order of $c/\omega_p$, which corresponds to $d \sim 100-500\;\mu$m for microplasma densities.    \par
The electron-ion collision frequency reads \begin{equation}
 	\nu_\text{ei} = \frac{Z^2 e^4 n_i \ln\Lambda_\text{ei}}{(4\pi\epsilon_0)^2m_e^{1/2} T_e^{3/2}} ,
 \end{equation}
where $n_i \approx n_e$ is the ion density and we take $Z=1$ for simplicity. The Coulomb logarithm is given by $\ln\Lambda_\text{ei} = \ln (b_{max}/b_{min})$ where $b_{min}$ and $b_{max}$ are, respectively, the distances of closest and farthest electron-ion approach, which we can estimate as $b_{min}= e^2/k_BT_e$ and $b_{max}$ equal to the Debye length. Using the parameters of table \ref{tab1}, we obtain $\ln\Lambda_\text{ei} \approx 30$, $\omega_p = 2 \pi \times 9  \times 10^{10} \ \text{Hz}$ and $\nu_\text{ei} = 5.6 \times 10^{8} \  \text{Hz}$. \par
\begin{figure}
\centering 
\includegraphics[scale=0.34]{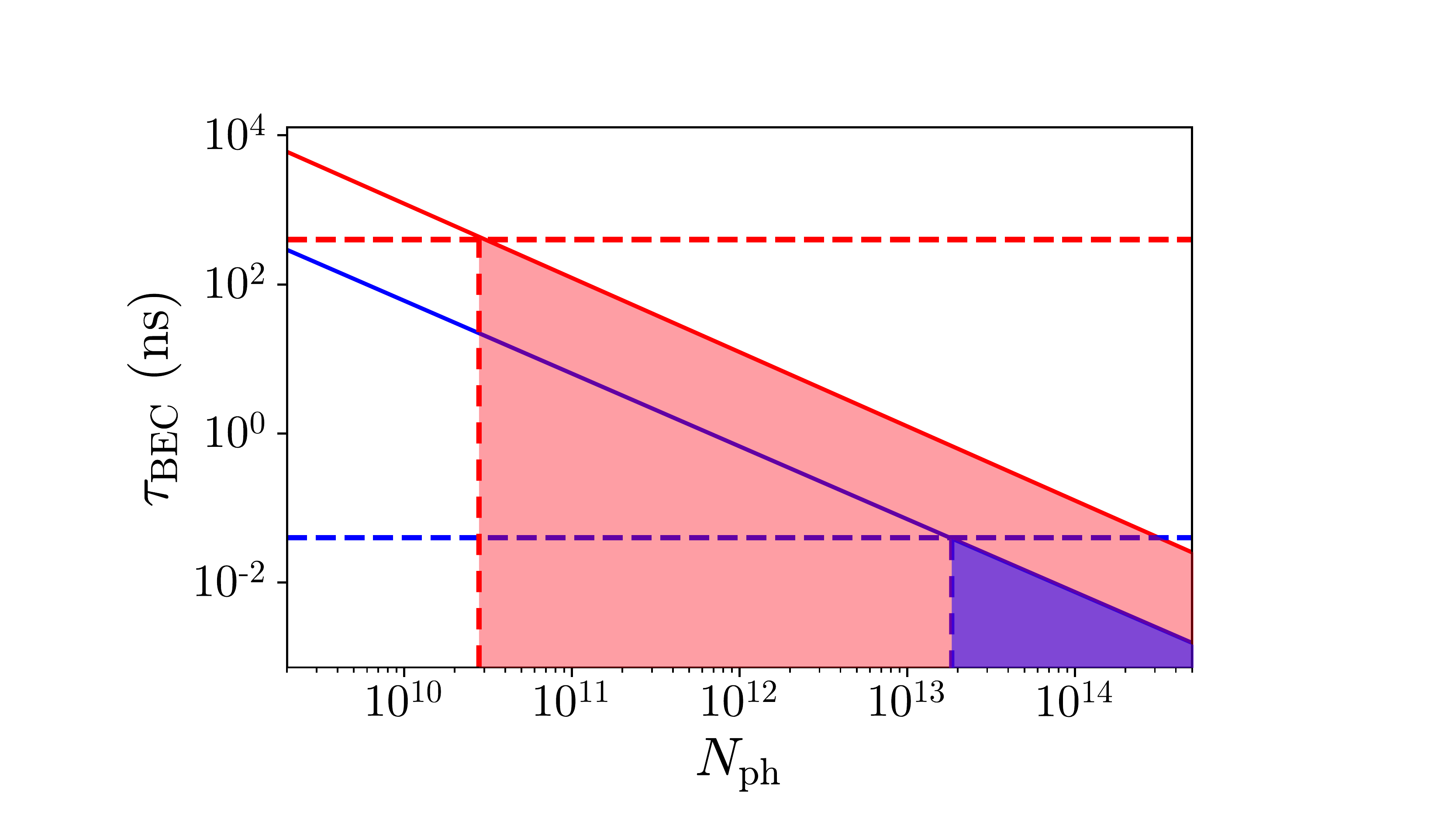}
\caption{Condensation time as a function of the number of photons for two different microplasmas (solid lines): $d=100\; \mu$m, $n_e = 10^{14}\; \text{cm}^{-3}$ and $T_e = 3\; $eV (red); $d=1\; $mm, $n_e = 10^{15}\; \text{cm}^{-3}$ and $T_e = 1\; $eV (blue). The red configuration is that of table ~\ref{tab1}. The dashed horizontal lines correspond to the maximum condensation time for each configuration before the effect of inverse bremsstrahlung becomes large. Shaded zones mark the regions where condensation is possible.}
\label{fig2}
\end{figure}

Equation \eqref{alpha} shows that the absorption coefficient is a growing function of $\kappa$, which in turn is maximum for minimum $\omega$. Therefore, if we estimate $\kappa$ for the lowest frequency, we obtain the maximum possible absorption coefficient valid for all photons inside the cavity. It results in $\alpha_\text{abs} \approx 6.2 \times 10^{-7}$. For the remaining photons, $\alpha_\text{abs}$ will attain even smaller values. The total absorption coefficient can now be estimated as 
\begin{equation}
	\alpha_\text{abs}^\text{tot}  = 1 - (1-\alpha_\text{abs})^{M}, \label{alphaTOT}
\end{equation}
where $M$ is the number of half round trips necessary for condensation. We obtain $M \sim 10^5$, which leads to
\begin{equation}
\alpha_\text{abs}^\text{tot} \sim 10^{-2} .
\end{equation}
Based on this value, we can deduce that absorption caused by inverse bremsstrahlung is expected to be insignificant for the desired conditions, therefore suggesting that it should be possible for the condensate to form. Reducing the number of photons would result in an increase of the condensation time, which would eventually lead to strong absorption during the time of the process. Having $N_\text{ph}$ of at least $10^{11}$ is thus crucial for the experimental implementation of photon condensation with the selected parameters (see figure~\ref{fig2}). 
